\documentclass[
reprint, 
superscriptaddress,
amsfonts, amsmath, amssymb,
aps,
pre,
titlepage,
]{revtex4-2}

\usepackage{graphicx}
\usepackage{dcolumn}
\usepackage{bm}
\usepackage{xcolor}
\usepackage{natbib}
\usepackage{cases}
\usepackage{amsmath}
\usepackage{mathtools}

\usepackage{hyperref}
\hypersetup{
    colorlinks=true,
    linkcolor=blue,
    filecolor=blue,      
    urlcolor=blue,
    allcolors=blue
    }
\usepackage[capitalize]{cleveref}
\usepackage{orcidlink}
\usepackage[mathlines]{lineno}
\usepackage[shortlabels,inline]{enumitem}

\definecolor{orange}{rgb}{1,0.5,0}
\definecolor{forestgreen}{rgb}{0.13, 0.55, 0.13}
\definecolor{bittersweet}{rgb}{1.0, 0.44, 0.37}
\definecolor{chartreuse}{rgb}{0.87, 1.0, 0.0}
\definecolor{darkorchid}{rgb}{0.6, 0.2, 0.8}

\usepackage[normalem]{ulem}

\usepackage[percent]{overpic}
\usepackage[]{xcolor}

\newcommand{\prob}{\mathbb{P}}
\newcommand{\qrob}{\mathbb{Q}}

\newcommand{\ms}{$\mathrm{m\,s}^{-1}$}

\DeclarePairedDelimiterX{\infdivx}[2]{(}{)}{%
  #1\;\delimsize\|\;#2%
}
\newcommand{\infdiv}{D_{KL}\infdivx}

\newcommand{\histogramswidth}{.92\linewidth}

\usepackage{subfiles}
\begin{document}

\preprint{APS/123-QED}

\title{Data-driven physics-based modeling of pedestrian dynamics}

\author{Caspar A.S. Pouw \orcidlink{0000-0002-3041-4533}}
\email[]{c.a.s.pouw@tue.nl}

\affiliation{Department of Applied Physics and Science Education, Eindhoven University of Technology, 5600 MB Eindhoven, The Netherlands}
\affiliation{ProRail BV, Moreelsepark 2, 3511EP Utrecht, The Netherlands}

\author{Geert van der Vleuten}
\affiliation{Department of Applied Physics and Science Education, Eindhoven University of Technology, 5600 MB Eindhoven, The Netherlands}

\author{Alessandro Corbetta \orcidlink{0000-0001-6979-3414}}
\affiliation{Department of Applied Physics and Science Education, Eindhoven University of Technology, 5600 MB Eindhoven, The Netherlands}
\affiliation{Eindhoven Artificial Intelligence Systems Institute, 5600 MB Eindhoven, The Netherlands}

\author{Federico Toschi \orcidlink{0000-0001-5935-2332}}
\affiliation{Department of Applied Physics and Science Education, Eindhoven University of Technology, 5600 MB Eindhoven, The Netherlands}
\affiliation{Eindhoven Artificial Intelligence Systems Institute, 5600 MB Eindhoven, The Netherlands}
\affiliation{CNR-IAC, I-00185 Rome, Italy}

\date{\today}

\begin{abstract}

Pedestrian crowds encompass a complex interplay of intentional movements aimed at reaching specific destinations, fluctuations due to personal and interpersonal variability, and interactions with each other and the environment. Previous work demonstrated the effectiveness of Langevin-like equations in capturing the statistical properties of pedestrian dynamics in simple settings, such as almost straight trajectories. However, modeling more complex dynamics, such as when multiple routes and origin-destinations are involved, remains a significant challenge.
In this work, we introduce a novel and generic framework to describe the dynamics of pedestrians in any geometric setting, significantly extending previous works. Our model is based on Langevin dynamics with two timescales. The fast timescale corresponds to the stochastic fluctuations present when a pedestrian is walking. The slow timescale is associated with the dynamics that a pedestrian plans to follow, thus a smoother path without stochastic fluctuations. Employing a data-driven approach inspired by statistical field theories, we learn the complex potentials directly from the data, namely a high-statistics database of real-life pedestrian trajectories. This approach makes the model generic as the potentials can be read from any trajectory data set and the underlying Langevin structure enables physics-based insights.
We validate our model through a comprehensive statistical analysis, comparing simulated trajectories with actual pedestrian measurements across five complementary settings of increasing complexity, including a real-life train platform scenario, underscoring its practical societal relevance. We show that our model effectively captures fluctuation statistics in pedestrian motion. 
Beyond providing fundamental insights and predictive capabilities in pedestrian dynamics, our model could be used to investigate generic active dynamics such as vehicular traffic and collective animal behavior.

\end{abstract}

\maketitle

\section{\label{sec:introduction}Introduction}
\begin{figure}
    \centering
    \includegraphics[width=\linewidth]{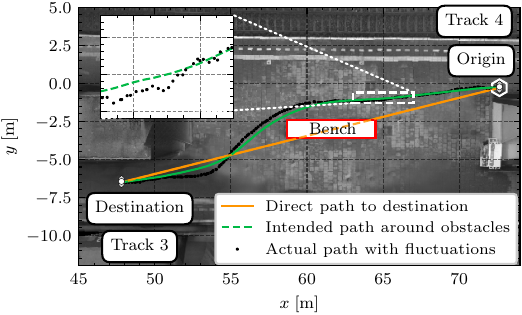}
    \caption{\label{fig:drawing-intended-path}Movement of a pedestrian on a platform in Eindhoven train station. The pedestrian walks from the origin, near track 4, to the destination, near track 3. A bench in the middle of the platform is marked with a red box. The destination is based on the train schedule, indicating strategic-level planning. The difference between the direct path (orange line) and planned path $\vec{x}_s(t)$ (green dashed line) illustrate tactical, obstacle-evading behavior around the bench. The actual path $\vec{x}(t)$ (black dots) represents the full dynamics including fluctuations induced by operational behavior. The inset highlights the fluctuations around the planned route.}
\end{figure}

Despite the perception that human behavior cannot be described by mathematical models, it is now well established that statistical descriptions can effectively represent the dynamics of individuals within crowds~\citep{hughes-arfm-2003, cristiani-book-2014, feliciani-book-2022, corbetta-annurev-2023}. It is widely accepted that pedestrian behavior takes place on multiple time scales, in the literature often categorized into three distinct levels~\citep{daamen-thesis-2004, chraibi-encyclopedia-2018}:
\begin{enumerate*}[label=(\roman*)]
\item long-term strategic behavior that determines an overall goal or objective; 
\item short-term tactical behavior associated with planning a smooth path toward that goal, possibly in dependence on a large number of parameters, such as travel time~\citep{kretz_jsm_2009,gabbana-pnasn-2022}, path length~\citep{hughes-arfm-2003}, energy expenditure~\citep{arechavaleta-ieee-2008}, road conditions~\citep{liang-be-2020, fossum-trpd-2021, miao-sustainability-2023}, and nearby individuals, walls, obstacles~\citep{chraibi-pre-2010, fujita-pre-2019, tong-jtrsi-2022};
\item immediate operational behavior such as lateral body fluctuations (i.e. sway~\citep{pauls-ped-2007, liu-physa-2009, parisi-pre-2016}) and collision avoidance, often through subconscious adjustments in response to the immediate environment~\citep{montello-book-2006, connor-jon-2012, selinger-cb-2015}. These fluctuations happen at a scale much shorter than, e.g., variations in planning.
\end{enumerate*}
We illustrate these behavioral levels based on a real-world pedestrian trajectory in Fig.~\ref{fig:drawing-intended-path}. Models shall thus incorporate a deterministic component, to account for strategic and tactical behavior, while operational aspects such as sway and quick readjustments have often been treated via a stochastic component, typically via additive noise, e.g.~\cite{corbetta-pre-2017,willems-scirep-2020,zanlungo-coldyn-2020}. The ability to fully describe these components and their interplay presents an outstanding scientific challenge, crucial to optimizing crowd simulation software used to design safer and more functional urban infrastructures.

Crowd-modeling techniques have been proposed at different scales, primarily categorized into macroscopic (Eulerian) and microscopic (Lagrangian) models~\citep{chraibi-encyclopedia-2018, corbetta-annurev-2023}. Macroscopic models address the crowd through an hydrodynamics perspective, thus averaging out stochastic fluctuations~\cite{cristiani-book-2014,bartolo-science-2019}. Here we consider microscopic models, that address the dynamics of single individuals, $\vec{x}(t)$, treating them as Newton-like active particles. 
These models employ molecular dynamics-like approaches, such as the Langevin framework, to capture the evolution of stochastic active particle systems, focusing on the interactions between agents based on ``ad hoc'' social forces~\citep{helbing-pre-1995, chraibi-pre-2010, cristiani-book-2014}. The Langevin framework leverages vector-valued stochastic differential equations to incorporate both deterministic forces, often derived from the gradient of a potential that could depend on position and velocity, and stochastic forces~\citep{langevin-1908}. This approach can be applied to a wide range of systems, including motile cells~\citep{shienbein-bmb-1993}, financial markets~\citep{bouchaud-epjb-1998}, and climate dynamics~\citep{berglund-sd-2002}. 

In pedestrian dynamics, previous work demonstrated that Langevin-like models can be tailored to quantitatively reproduce observations in prototypical environments such as narrow corridors~\citep{corbetta-pre-2017}, curved paths~\citep{vleuten-pre-2024}, simple pairwise avoidance~\citep{corbetta-pre-2018} and small social groups~\cite{zanlungo-pre-2014}. Yet, traditional Langevin-like approaches, as in~\citep{corbetta-pre-2017,corbetta-pre-2018}, rely on assumptions of geometric confinement, homogeneity, and symmetries in space and velocity for the potentials. These assumptions unavoidably fail as we consider realistic geometric settings where the effect of confinement by the environment (e.g. due to a narrow corridor) disappears. According to their planning (tactical scale), pedestrians exhibit trajectories that are often straight regardless the confinement of a corridor (cf. Fig.~\ref{fig:drawing-intended-path}), thus e.g. in wide areas, around cross-points or intersections. The richer symmetry structure of these settings requires a description with further degrees of freedom encompassing the tactical dynamics. For instance, a wide corridor has a continuous symmetry group of translations in the direction transversal to the corridor, absent in a narrow corridor. In technical terms, these require a more generic potential entailing tactical aspects, e.g. via additional dimensions in space and velocity, and that thus break these symmetries.  

Here, our main goal is to create a single model that can be applied to any environment - as such, we aim to generalize  existing models~\citep{corbetta-pre-2017,corbetta-pre-2018,vleuten-pre-2024} and to extend to environments not described before (e.g. intersecting paths). We present a data-driven, physics-based approach that enables us to leverage the Langevin framework to develop a generalized pedestrian model applicable to any environment. Inspired by the methods described in~\citep{frishman-prex-2020}, we derive an empiric potential from trajectory recordings to compute deterministic forces while treating stochastic forces as noise. Our goal is to create a method that can be easily used to generate trajectories closely matching the statistical properties of real pedestrians. A model of this kind provides fundamental insights in the kinematics of pedestrians, and aids urban design to optimize safety and efficient crowd flow. Our Python implementation is open-source on GitHub~\cite{pouw-physped-2024}, providing a tool for extracting physical insights about the dynamics of pedestrians and their environment, and facilitating simplified approximations of these dynamics.

Generalizing~\citep{corbetta-pre-2017, corbetta-pre-2018}, we describe the motion of a single individual (1-pedestrian model) with the following Langevin equation
\begin{equation}\label{eq:langevin-model}
    \begin{dcases}
        \dot{\vec{x}}(t) &= \vec{u}(t), \\
        \dot{\vec{u}}(t) &= -\left[\frac{\partial}{\partial \vec{x}} + \frac{\partial}{\partial \vec{u}}\right] U\left(\vec{x}(t), \vec{u}(t) \mid \begin{array}{c} \text{\scriptsize{tactical}} \\[-0.5em] \text{\scriptsize{aspects}} \end{array}\right) + \sigma \dot{\vec{W}},
    \end{dcases}
\end{equation} 
with the position $\vec{x}(t) = (x(t),y(t))$ and the velocity $\vec{u}(t)=(u(t),v(t))$ of the individual as a function of time, the deterministic force described by the gradient - denoted with $\left[\frac{\partial}{\partial \vec{x}} + \frac{\partial}{\partial \vec{u}}\right]$ - of a potential $U\left(\vec{x}, \vec{u} \mid \begin{array}{c} \text{\scriptsize{tactical}} \\[-0.6em] \text{\scriptsize{aspects}} \end{array}\right)$ that depends on the pedestrians' position $\vec{x}(t)$, velocity $\vec{u}(t)$, and tactical behavior, and the stochastic fluctuations modeled as an isotropic $\delta$-correlated in-time Gaussian noise $\dot{\vec{W}}$ with variance $\sigma^2$. These assumptions on the noise, while not required, are commonly used in the field and validated in many contexts~\citep{corbetta-pre-2017, corbetta-pre-2018, romanczuk-epj-2012}. 

We decompose the dynamics of a pedestrian into a slow and a fast component, with the slow element, $\vec{x}_s(t)$, representing tactical behavior (planning a smooth path) as a hidden variable on a slow manifold, and the fast element describing the operational behavior (body sway and collision avoidance), influenced by the stochastic noise. The full dynamics can be rewritten as superimposing the fast dynamics to the slow manifold. On these bases, we write Eq.~\eqref{eq:langevin-model} as
\begin{equation}\label{eq:generalized-model}
    \begin{dcases}
        \dot{\vec{x}}(t) &= \vec{u}(t)\\
        \dot{\vec{u}}(t) &= -\left[\frac{\partial}{\partial \vec{x}} + \frac{\partial}{\partial \vec{u}}\right]U\left(\vec{x}(t), \vec{u}(t) \mid \vec{x}_s(t), \vec{u}_s(t)\right) + \sigma \dot{\vec{W}}, \\
    \end{dcases}
\end{equation}
with the potential $U\left(\vec{x}(t), \vec{u}(t) \mid \vec{x}_s(t), \vec{u}_s(t)\right)$, now conditioned on the slow position, $\vec{x}_s$, and velocity, $\vec{u}_s$. This allows us to compute the deterministic forces associated with a pedestrian's tactical behavior. In the model Eq.~\eqref{eq:generalized-model} we have assumed that the tactical aspects in Eq.~\eqref{eq:langevin-model} are only represted by $\vec{x}_s$ and $\vec{u}_s$.

In this framework, pedestrian trajectories, described by the dynamics detailed in Eq.~\eqref{eq:generalized-model}, form the basis for defining this potential. This enables us to infer the potential from existing trajectory datasets. For instance, by utilizing public datasets from various real-world measurements~\citep{corbetta-4tu-2017, gabbana-zenodo-2022, heuvel-4tu-2022, pouw-cd-2021} and experimental environments~\citep{PedestrianDynamicsDataArchive}. 

Our data-driven approach connects the probability of a state $(\vec{x}, \vec{u} \mid \vec{x}_s, \vec{u}_s)$ to the potential associated with that state, $U(\vec{x}, \vec{u} \mid \vec{x}_s, \vec{u}_s)$, via the Gibbs measure i.e.
\begin{equation}\label{eq:gibbs-measure}
    \prob(\vec{x}, \vec{u} \mid \vec{x}_s, \vec{u}_s) \propto e^{-\,U(\vec{x}, \vec{u} \mid \vec{x}_s, \vec{u}_s)}.
\end{equation}
It must be noted that this approach also enables us to simulate the trajectory of a single pedestrian in a dense crowd. In that situation, the potential $U\left(\vec{x}(t), \vec{u}(t) \mid \vec{x}_s(t), \vec{u}_s(t)\right)$ in Eq.~\eqref{eq:generalized-model} should be interpreted as an {\em effective} potential, renormalized by the mutual interactions within the crowd. Possibly, $U$ may even carry a time dependence as the level of crowdedness may change over time.

This paper is structured as follows. In Sec.~\ref{sec:pedestrian-dynamics-in-a-corridor} we re-express previous works on specific Langevin models for pedestrian dynamics in corridors in terms of the generalized pedestrian model in Eq.~\eqref{eq:generalized-model}. In Sec.~\ref{sec:generalized-pedestrian-model} we expand on the generalized model by closing the dynamical system, and introducing a data-driven method to extract the potential from pedestrian measurements. In Sec.~\ref{sec:model-validation} we validate our model by testing it in different environments. In Sec.~\ref{sec:path-integral} we briefly discuss our methods and results in the context of the path-integral framework. We conclude with a discussion and outlook in Sec.~\ref{sec:conclusion}.
\section{\label{sec:pedestrian-dynamics-in-a-corridor}Modeling pedestrians in a corridor}
In this section, we re-express two specific Langevin-based pedestrian models, from previous work by some of the authors, in terms of the generalized pedestrian model in Eq.~\eqref{eq:generalized-model} and study the associated potentials $U(\vec{x}, \vec{u} \mid \vec{x}_s, \vec{u}_s)$. Describing these simple settings with our generalized model aids in understanding the methods in more complex settings. Respectively, we consider the models in~\citep{corbetta-pre-2017}, addressing single pedestrian movement through a narrow corridor, and in~\citep{corbetta-pre-2018},  focusing on a wide corridor with multiple parallel paths. 
\begin{figure}
    \centering
    \begin{overpic}[width=\linewidth]{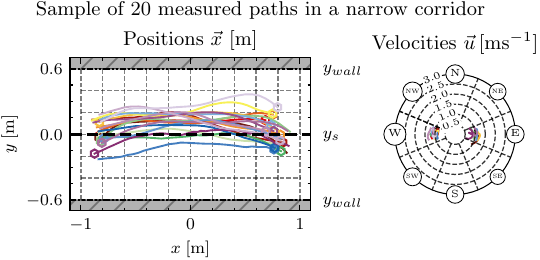}
        \put(0, 39){(a)}
        \put(60, 39){(b)}
    \end{overpic}
    \begin{overpic}[width=\linewidth]{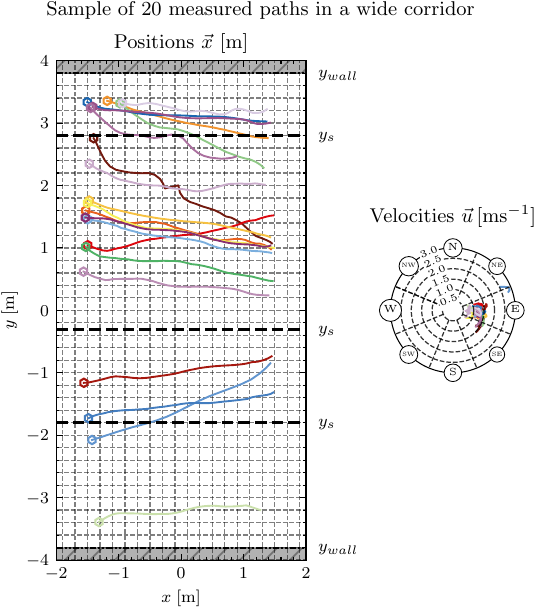}
        \put(0, 83){(c)}
        \put(60, 83){(d)}
    \end{overpic}
    \caption{\label{fig:corridor-trajectories}Pedestrian trajectories measured inside two corridors. For clarity, selections comprising only 20 trajectories are shown. (a, b) Positions, $\vec{x}(t)$, (a) and velocities, $\vec{u}(t)$, (b) over time at the Metaforum building at Eindhoven University of Technology. Slow positions, $y_s(t)=0$ shown by the black dashed line. (c, d) Positions, $\vec{x}(t)$, (c) and velocities, $\vec{u}(t)$, (d) over time in a wide corridor at Eindhoven Central train station. Three possible realizations of slow positions $y_s(t)$ are visualized with black dashed lines, these would represent pedestrians walking through the corridor at different distances, $y$, from the walls.
    }
\end{figure}

\subsection{\label{ssec:narrow-corridor}Walking through a narrow corridor}
The dynamics of a single pedestrian walking undisturbed through a narrow corridor is possibly one of the simplest conceivable dynamics; an almost one-dimensional movement solely confined by two parallel corridor walls~\citep{corbetta-pre-2017}. The authors use a coordinate system $(x,y)$ where $x$ aligns with the direction of the corridor. As such, the corridor is spatially homogeneous along the x-axis with $y = 0$ representing the corridor's longitudinal axis. In Fig.~\ref{fig:corridor-trajectories}(a, b), we report a selection with 20 measured trajectories.

The simplicity of the geometry and the confinement due to the environment allowed the authors two simplifications: decomposing the potential $U(\vec{x}, \vec{u} \mid \vec{x}_s, \vec{u}_s)$ into a position dependent term $U(\vec{x} \mid \vec{x}_s)$ and a velocity dependent term $U(\vec{u} \mid  \vec{u}_s)$, and assuming that the longitudinal and transverse dynamics are uncorrelated. As such, the potential can be written as
\begin{align}\label{eq:factorized-potential}
\begin{split}
    U(\vec{x}, \vec{u} \mid \vec{x}_s, \vec{u}_s) = \underbrace{U(x \mid x_s) + U(y  \mid y_s)}_{U(\vec{x} \mid \vec{x}_s)} \\ +\underbrace{U(u \mid u_s) + U(v \mid v_s)}_{U(\vec{u} \mid \vec{u}_s)}.
\end{split}
\end{align}
Note that here, differently from e.g. \cite{corbetta-pre-2017} and \cite{corbetta-pre-2018}, we use the symbol $U$ to jointly represent both position and velocity potentials. The symmetry of the corridor also allowed the authors to formulate reasonable assumptions about the tactical behavior of pedestrians, thereby describing their slow dynamics. They assume pedestrians form straight trajectories parallel to and equidistant from both walls, aligned with the center of the corridor $y_s(t)=0$, walking at an average walking pace, $\langle v \rangle \approx 1.3$~\ms~\citep{buchmueller-parameters-2006, feliciani-plos-2018} in either direction through the corridor, thus $u_{s}(t) = \pm 1.3$~\ms and $v_s(t)=0$~\ms.

\begin{figure}[ht]
    \centering
    \begin{overpic}[width = \linewidth]{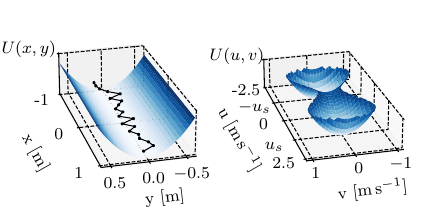}
        \put(0, 42){(a)}
        \put(45, 42){(b)}
    \end{overpic}
    \begin{overpic}[width=\linewidth, trim=-.5cm 0.5cm -.5cm 1cm, clip]{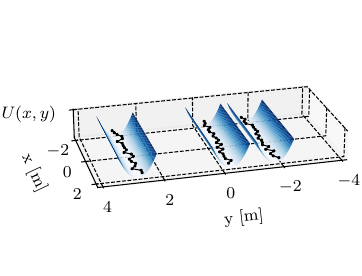}
        \put(10, 38){(c)}
    \end{overpic}
    \caption{\label{fig:potentials-narrow-corridor}Three-dimensional representations of the potentials within the corridors. (a) Position confinement potential $U(x,y)$ in a narrow corridor (cf. Eq.~\eqref{eq:space-potential-narrow-corridor}) centered in $y_s(t)=0$~m.
    (b) Velocity potential $U(u,v)$ (cf. Eq.~\eqref{eq:velocity-potential-narrow-corridor}) with preferred walking velocity $u_s(t) \approx \pm1.3$~\ms. (c) Confinement potential in the wide corridor, showing three different trajectories parallel to the walls but at different distances.
    }
\end{figure}

Based on these assumptions, the authors derived potentials that describe the governing dynamics. The longitudinal movement along $y_s(t) = 0$, indicating no confinement in that direction, results in a flat potential. The lateral movement, confined by the corridor walls, requires a restoring force towards the center of the corridor, $y_s(t) = 0$, modeled with a parabolic potential centered around $y=0$ (a simplifying assumption supported by the data). We illustrate the potential in Fig.~\ref{fig:potentials-narrow-corridor}a. In formulas, they describe the confinement potential as
\begin{equation}\label{eq:space-potential-narrow-corridor}
    \begin{cases}
        U(x \mid x_s) &= c \\
        U(y \mid y_s) &= \beta \, y^2,
    \end{cases}
\end{equation}
with $c$ a constant and $\beta$ the curvature of the parabolic potential. The velocity in the transverse direction, $v_s(t)=0$~\ms, is also modeled with a parabolic potential centered around $v=0$~\ms. The velocity component along the longitudinal direction, $u_{s}(t) = \pm 1.3$~\ms, is modeled with a double-well potential with curvature~$\alpha$; possibly the lowest-order model that displays the presence of bi-stable states. This reads
\begin{equation}\label{eq:velocity-potential-narrow-corridor}
    \begin{cases}
    U(u \mid u_s) &= \alpha \, (u^2 - u^2_s)^2 \\
    U(v \mid v_s) &=  \nu \, v^2,
    \end{cases}
\end{equation}
with $\alpha$ and $\nu$ the curvature in the longitudinal and transverse direction respectively. In Fig.~\ref{fig:potentials-narrow-corridor}b we report a 3-dimensional representation of the longitudinal velocity potential for $v=0$~\ms. 

\subsection{\label{ssec:wide-corridor}Walking through a wide corridor}
Movement through any corridor is primarily confined (or guided) by the presence of the walls, directing the flow in the longitudinal direction. Consequently, the authors describe the slow velocities again with $u_s(t)=1.3$~\ms~and $v_s(t)=0$~\ms. Paths on the slow manifold develop as quasi-rectilinear trajectories in this direction. However, in contrast to the narrow corridor, the wide corridor enables a continuous choice of straight paths. In Fig.~\ref{fig:corridor-trajectories}(c,d) we report a selection of 20 trajectories measured within a wide corridor. The data show straight paths traversing the corridor following a quasi-straight continuation of their initial $y$ coordinate. In the figure, we illustrate three different possible realizations of slow positions $y_s(t)$ to show the dependence on the initial position of the pedestrian. In Fig.~\ref{fig:potentials-narrow-corridor}c we show three parabolic confinement potentials, each for a different initial position $y_s(t) = y(0)$. For an extension to the wide corridor model that includes pedestrian interactions, we refer to previous work by some of the authors, i.e., for the case $1\, \mathrm{vs.}\, 1$ cf.~\citep{corbetta-pre-2018}, and for the case $1\, \mathrm{vs.}\, N$ cf.~\citep{corbetta-crowddyn-2020}.

\subsection{\label{ssec:corridor-limitations}Limitations of the corridor model}
The challenges of employing the generalized model in Eq.~\ref{eq:generalized-model} to simulate any dynamics are two-fold: on one hand, it is not possible to decouple the potential into separate components, resulting in a complex multidimensional potential; and, on the other hand, the slow dynamics are unknown. Solutions to these challenges will be discussed in the next section.
\begin{figure*}
    \begin{overpic}[width = \columnwidth, trim = 0 -50 0 0]{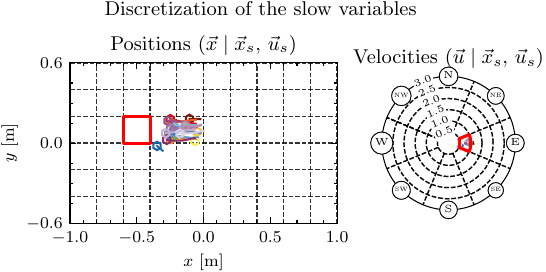}
        \put(0, 62){(a)}
        \put(67, 62){(b)}
    \end{overpic}
    \begin{overpic}[width = \columnwidth]{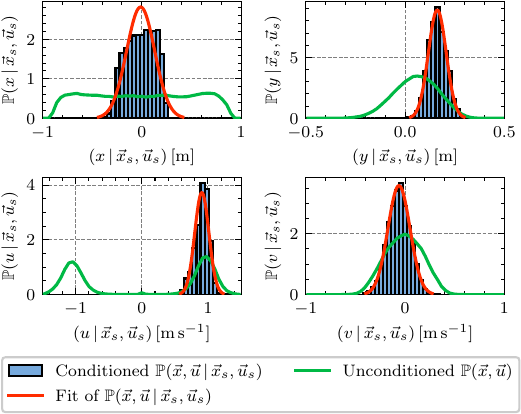}
        \put(0, 79){(c)}
        \put(50, 79){(d)}
        \put(0, 45){(e)}
        \put(50, 45){(f)}
    \end{overpic}
    \caption{\label{fig:fast-and-slow-trajectories} Potential approximation in the narrow corridor. (a, b) Representation of the discretization of  slow dynamics in terms of positions $\vec{x}_s$ (a) and velocities $\vec{u}_s$ (b). One lattice cell (red border) encloses the point $x_s \approx -0.4$ m, $y_s \approx 0.1$ m, $u_{r,s} \approx 1.2$, and $u_{\theta,s} \approx 0$ deg. This cell captures the trajectories of pedestrians walking in eastern direction. We display a subset with 20 trajectories conditioned to the highlighted lattice cell. (c--f) Probability distributions of longitudinal positions $x$ (c), transversal positions $y$ (d), longitudinal velocities $u$ (e), and transversal velocities $v$ (f) conditioned on the lattice cell indicated in panels (a,b). The distributions are fitted with a second-order polynomial (shown in red) to locally approximate the potential, $U(\vec{x},\vec{u}\mid\vec{x}_s,\vec{u}_s)$. The green line represents the unconditioned distributions. All distributions are normalized to an area of 1.
    }
\end{figure*}
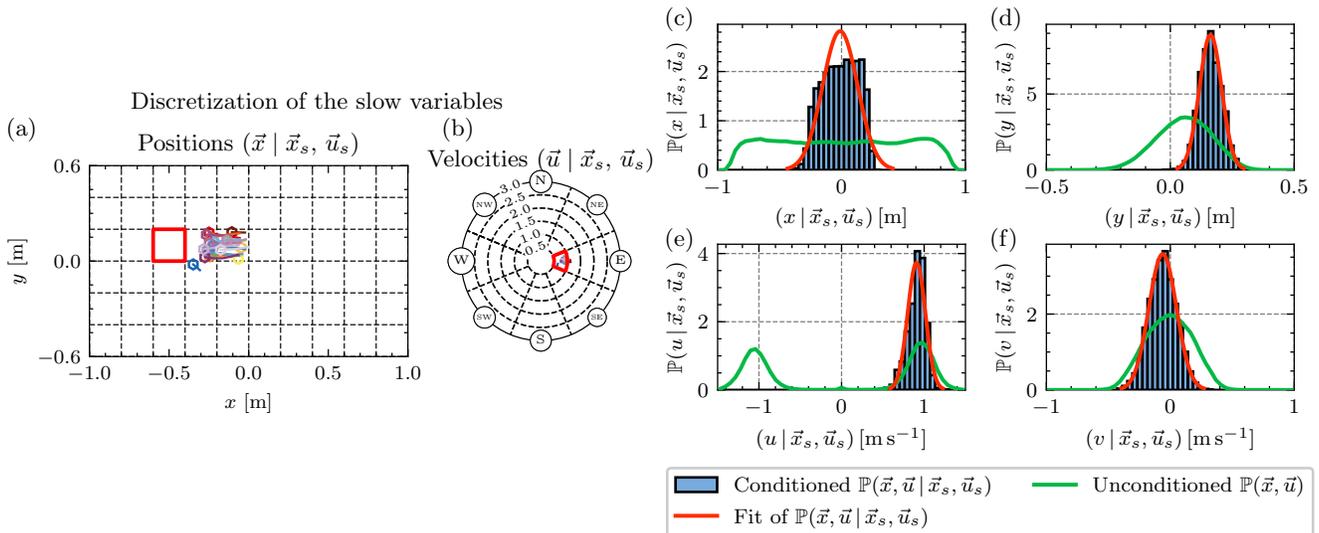
    
\section{\label{sec:generalized-pedestrian-model}Generalized pedestrian model}
In this section, we expand on the generalized pedestrian model from Eq.~\eqref{eq:generalized-model}. We close the dynamical system of Eq.~\eqref{eq:generalized-model} with mathematical expressions to approximate the slow modes $\vec{x}_s$ and $\vec{u}_s$. We present a data-driven approach to extract the high-dimensional potential $U(\vec{x},\vec{u}\mid\vec{x}_s,\vec{u}_s)$ from large-scale, pedestrian trajectory data sets. Finally, we provide guidelines on how to calibrate the model parameters.

\subsection{Slow dynamics}
The slow dynamics, $\vec{x}_s$ and $\vec{u}_s$, represent tactical behavior (planning a smooth path) as hidden variables on a slow manifold. While in the case of the narrow corridor one can make a reasonable guess a priori for the values of the slow variables ($y_s=0$), in general a relevant issue to close Eq.~\eqref{eq:generalized-model} is how to estimate the slow dynamics ($\vec{x}_s$, $\vec{u}_s$) from the data themselves.  We derive the slow dynamics by filtering out high-frequency fluctuations from the actual dynamics; practically, we define the slow derivatives $\dot{\vec{x}}_s(t)$ and $\dot{\vec{u}}_s(t)$ as first-order low-pass filters of the actual positions $\dot{\vec{x}}$ and velocities $\dot{\vec{u}}(t)$, with a relaxation time $\tau$. This gives the following closed, self-consistent, system of differential equations

\begin{subnumcases}{\label{eq:generalized-model-with-closure}}
    \dot{\vec{x}}(t) = \vec{u}(t) \label{eq:model-with-closure-a} \\
    \dot{\vec{u}}(t) = -\left[\frac{\partial}{\partial \vec{x}} + \frac{\partial}{\partial \vec{u}}\right]U\left(\vec{x}, \vec{u} \mid \vec{x}_s, \vec{u}_s\right) + \sigma \dot{\vec{W}} \label{eq:model-with-closure-b} \\
    \dot{\vec{x}}_s(t) = -\frac{1}{\tau} \left(\vec{x}_s(t) - \vec{x}(t)\right) \label{eq:model-with-closure-c} \\
    \dot{\vec{u}}_s(t) = -\frac{1}{\tau} \left(\vec{u}_s(t) - \vec{u}(t)\right) \label{eq:model-with-closure-d}.
\end{subnumcases}

Thus, the low-pass filters describe a resistance to perturbations in the dynamics, thereby acting as an inertia, with characteristic time scale $\tau$. This also introduces a small delay in the dynamics of the slow variables with respect to the full dynamics.

\subsection{\label{ssec:data-driven-potential-approximation}Data-driven potential}
We approximate an empiric Langevin potential using the Gibbs measure from Eq.~\eqref{eq:gibbs-measure}. Our approximation leverages on the availability of vast ensembles of $N$ trajectories $\{\vec{x}^1(t),\vec{x}^2(t),\vec{x}^3(t),\ldots ,\, \vec{x}^N(t)\}$, and hinges on the following four steps:

\begin{enumerate}
\item We built the lifted representation of each trajectory of the ensemble
\begin{equation}
    \vec{x}^i(t) \mapsto \left(\,\vec{x}^i(t), \vec{u}^i(t), \vec{x}_s^i(t), \vec{u}^i_s(t)\,\right)\ \quad \forall i,   
\end{equation}
where $\vec{u}^i(t)$, $\vec{x}_s^i(t)$, and $\vec{u}^i_s(t)$ are computed, respectively, via Eq.~\eqref{eq:model-with-closure-a}, Eq.~\eqref{eq:model-with-closure-c}, and Eq.~\eqref{eq:model-with-closure-d}. This enables us to determine the empiric probability of the state $(\vec{x}, \vec{u})$ as a simple histogram
\begin{equation}
\{\vec{x}^i(t), i=1,\ldots\}\ \rightarrow\ \prob(\vec{x}, \vec{u}),
\end{equation}
as well as its counterpart conditioned to the slow variables $(\vec{x}_s, \vec{u}_s)$
\begin{equation}
\{\vec{x}^i(t), i=1,\ldots\}\ \rightarrow\ \prob(\vec{x}, \vec{u} \mid \vec{x}_s, \vec{u}_s).
\end{equation}

\item 
We employ the relation in Eq.~\eqref{eq:gibbs-measure}, yet after a factorization of the joint conditional probability $\prob (\vec{x}, \vec{u} \mid \vec{x}_s, \vec{u}_s)$ into the product of the marginal conditional probabilities (this step, which reduces the generality of $U$, is only aimed at reducing the computational complexity), i.e.
\begin{align}
    U(\vec{x}, \vec{u} \mid \vec{x}_s, \vec{u}_s) &\approx - A \ln \left[\prod_{z\in\{x,y,u,v\}}  \prob (z \mid \vec{x}_s, \vec{u}_s)\right] \\
    & = - A \sum_{z\in \{x,y,u,v\}} \ln \left[\prob (z \mid \vec{x}_s, \vec{u}_s) \right] \label{eq:emperic-potential}\\
    & = - A \sum_{z\in \{x,y,u,v\}} U(z \mid \vec{x}_s, \vec{u}_s) ,
\end{align}
where $A$ is a normalization constant. 
\item 
we approximate the marginals, $\prob (z \mid \vec{x}_s, \vec{u}_s), z\in\{x,y,u,v\}$, via Gaussian distributions. These local approximation are parametrized on the slow variables, $x_s, y_s, u_s, v_s$. For example, for the position $x$, the approximation reads 
\begin{equation}
    \prob (x \mid \, \vec{x}_s,\vec{u}_s) \sim \mathcal{N}\left(\mu_x \left(\vec{x}_s,\vec{u}_s\right), \xi_x(\vec{x}_s,\vec{u}_s) \right),
\end{equation}
where $\mu_x \left(\vec{x}_s,\vec{u}_s\right)$ and $\xi_x \left(\vec{x}_s,\vec{u}_s\right)$ are the mean and standard deviation of the local Gaussian approximation. Fig.~\ref{fig:fast-and-slow-trajectories}b reports the local approximation of the marginals inside a narrow corridor. It is worth noting that this Gaussian approximation is not strictly necessary, and more complex approximations can be easily employed. The assumption of Gaussian shape will have to be supported by the data.

\item
Effectively, we work with the following $(\vec{x}_s,\vec{u}_s)$-dependent second order polynomial approximation
that, e.g. for the $x$ potential,  reads
\begin{equation}
    U(x \mid \vec{x}_s, \vec{u}_s) = \beta_x(\vec{x}_s,\vec{u}_s) (x - \mu_x(\vec{x}_s,\vec{u}_s))^2,
\end{equation}
where the leading order coefficient (characterizing the curvature of the parabola, i.e. the stiffness of the potential) satisfies 
\begin{equation}
\beta_x(\vec{x}_s,\vec{u}_s) = \frac{A_x(\vec{x}_s,\vec{u}_s)}{2\, \xi_{x}(\vec{x}_s,\vec{u}_s)^2}.
\end{equation} 
We write the potentials along $y,u,v$ in a similar fashion and in terms of $\mu_y$, $\mu_u$, $\mu_v$ and $\xi_{y}, \xi_{u}$, $\xi_{v}$, respectively. The scaling factors satisfy $A_x(\vec{x}_s,\vec{u}_s) = \xi_u(\vec{x}_s,\vec{u}_s)^2$, $A_y(\vec{x}_s,\vec{u}_s)=\xi_v(\vec{x}_s,\vec{u}_s)^2$, and $A_u=A_v=\sigma^2/2$. These coefficients can be derived by considering stationary solutions of the Fokker-Planck equation~\cite{risken-book-1996} associated with the $(\vec{x},\vec{u})$ physical dynamics (cf. Appendix \ref{app:scaling-factors}).
\end{enumerate}
Note that in this procedure the noise intensity and relaxation time, i.e. $(\sigma,\tau)$, come as  parameters.

\subsection{\label{ssec:model-parameters}Lattice discretization}
We discretize the slow dynamics $(\vec{x}_s, \vec{u}_s)$ on a lattice-based grid, $\Lambda$, in such a way that the free coefficients  of the potential (e.g. $\mu_x(\vec{x}_s, \vec{u}_s)$, $\beta_x(\vec{x}_s, \vec{u}_s)$, etc.) are constant in each grid cell. We construct our lattice as the product 
\begin{equation}
    \Lambda = \Lambda_{xy} \times \Lambda_{v_r v_{\theta}}
\end{equation} 
with $\Lambda_{xy}$ a Cartesian lattice discretizing slow positions, $(x_s, y_s)$, and $\Lambda_{v_r v_{\theta}}$ a lattice in polar coordinates discretizing the slow velocities $(u_s,v_s)$. Describing the velocities in terms of direction and magnitudes enables us to discretize efficiently velocities within a given magnitude. The estimation procedure in Sec.~\ref{ssec:data-driven-potential-approximation}, in dependence on $(\sigma,\tau)$, allows us to determine the values of $\mu$ and $\xi$  for all components ($x,y,u,v$), for all lattice cells in $\Lambda$. 

Let $\prob_j$ be the density of the empiric probability in Eq.(10) based on the binning defined by the lattice $\Lambda$. In other terms, $\prob_j$ is the values of the normalized histogram in lattice cell $j$ and 
\begin{equation}
    \sum_j \prob_j \Delta (\Lambda_j) = 1
\end{equation}
holds. Note that the volumes of the lattice cells satisfy 
\begin{equation}
 \Delta (\Lambda_j) = \Delta x_j \cdot \Delta y_j \cdot \Delta u_{r,j} \cdot  |\vec {u}_j| \Delta u_{\theta,j},
\end{equation}
and are not uniform due to the polar discretization. Similarly, we can determine $\qrob_i$ based on repeated simulation of Eq.~\eqref{eq:generalized-model-with-closure}.

The goal is to have faithful simulation of the dynamics, i.e. we aim at $\qrob_i \approx \prob_i$. We retain the commonly used Kullback-Leibler divergence (KL, or relative entropy~\citep{kullback-ams-1951}) to quantify the difference between probabilities
\begin{equation}\label{eq:kl-divergence}
\infdiv{\prob}{\qrob} = \sum_{j} \prob_j \log{\left(\frac{\prob_j}{\qrob_j}\right)}\, \Delta(\Lambda_j).
\end{equation}
\section{\label{sec:model-validation}Results - Validation}
In this section we validate our model, Eq.~\eqref{eq:generalized-model-with-closure}, employing five increasingly complex scenarios (A-E, see Table~\ref{tab:validations}). We show how it not only generalizes existing works~\citep{corbetta-pre-2017, corbetta-pre-2018, vleuten-pre-2024}, but also allows to treat cases that are challenging due to lack of confinement and/or due to the presence of intersecting paths (cf. Sec.~\ref{sec:introduction}). For each test we set the noise intensity $\sigma = 0.9$ m s$^{-3/2}$ and the relaxation time $\tau = 0.5$ s. Thereby demonstrating the model’s general applicability as we do not need to fine-tune the $(\sigma, \tau)$-pairs to the specific scenarios (though this may be possible). In a numerical perspective, for each test we take the following steps:
\begin{enumerate}
    \item we construct the lattice $\Lambda$ using spatial cells with dimensions of $\Delta x = \Delta y \simeq 0.2$~m, about half the typical size of a human. The velocity lattice, $\Lambda_{u_r u_{\theta}}$, employs five radial bins $\Delta u_r = 0.5$~\ms~and eight angular segments $\Delta u_{\theta} = 45^{\circ}$ (along the principal compass directions). Velocities lower than $v_r<0.5$~\ms~are not categorized by direction, as we consider the direction irrelevant at low walking speeds. (cf. Fig~\ref{fig:fast-and-slow-trajectories}a).
    \item we establish the free parameters of the potential $U(\vec{x}, \vec{y} \mid \vec{x}_s, \vec{u}_s)$ (i.e. $\mu_x$, $\beta_x$ and so on) using the measurements following the procedure in Sec.~\ref{sec:generalized-pedestrian-model};
    \item we simulate trajectories by numerically integrating Eq.~\eqref{eq:generalized-model-with-closure} (cf. Appendix~\ref{app:numerical-simulations}). We use a timestep, $\Delta t$, inversely proportional to the sampling frequency of the measurements, $\Delta t = 1/f$;
   \item we compare probability distributions in terms of the Kullback-Leibler (KL) divergence Eq.~\eqref{eq:kl-divergence};
    \item optional: we could run an optimization process to fine-tune the noise intensity, $\sigma$, and the relaxation time, $\tau$, that minimize the Kullback-Leibler divergence $\infdiv{m}{s}$. 
\end{enumerate}

\begin{figure*}
    \begin{overpic}[width=\columnwidth, trim = 0 -10 0 0]{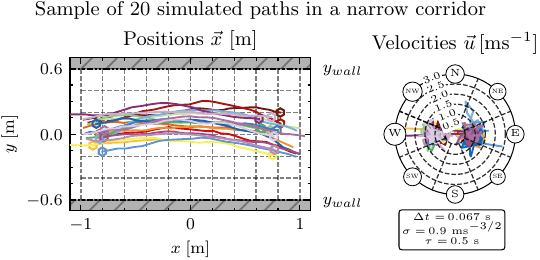}
        \put(-5,50){\Large A)}
        \put(0, 43){(a)}
        \put(60, 43){(b)}
    \end{overpic}
    \begin{overpic}[width=.9\columnwidth, trim = 0 -10 0 0]{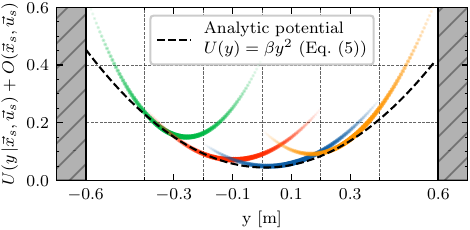}
        \put(0,56){(c)}
    \end{overpic}
    \begin{overpic}[width=\histogramswidth]{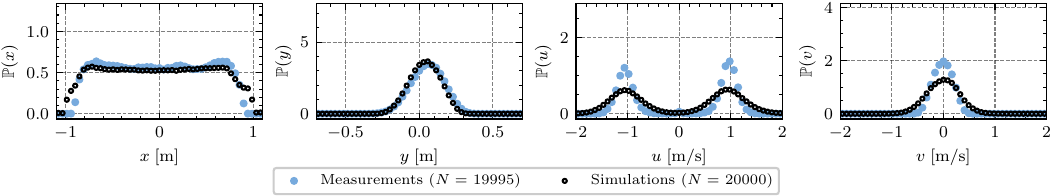}
        \put(0, 18){(d)}
        \put(26, 18){(e)}
        \put(51, 18){(f)}
        \put(76, 18){(g)}
    \end{overpic}
    \caption{\label{fig:narrow-corridor-trajectories} Modeling single straight walking paths. (a, b) A selection of 20 synthetic trajectories (model parameters provided in the inset). The synthetic trajectories closely match the real-world trajectories, see Fig.~\ref{fig:corridor-trajectories}a. (c) Cross-section of the data-driven potential, $U(y \mid \vec{x}_s, \vec{u}_s)$, along the $x=0$ axis of the corridor, and conditioned to slow variables $x_s \approx 0$~m, $u_s \approx 0.7$~\ms, and $v_s \approx 0$~\ms. Each colored line segment is conditioned to a different slow y-position, with $y_s=\{-0.3, -0.1, 0.1, 0.3\}$. An offset $O(\vec{x}_s, \vec{u}_s) = A \ln{[\mathbb{P}(\vec{x}_s, \vec{u}_s)}]$ is added for visualization, this does not influence the dynamics (Eq.~\eqref{eq:generalized-model-with-closure}) which is governed only by free parameters mean, $\mu_y(\vec{x}_s, \vec{u}_s)$, and curvature, $\beta_y(\vec{x}_s, \vec{u}_s)$. The curvature of the potential pieces is in good agreement with the analytical potential (black dashed line), shown as $U(y) = \beta y^2$ with $\beta = 1.9$ (cf. Eq.~\eqref{eq:space-potential-narrow-corridor}).   Note that the analytic potential is a simplification, and the data-driven potential is a better approximation as already shown in \cite{corbetta-pre-2017}. (d--g) Comparative analysis of probability distributions between simulated (open dots) and recorded (blue dots) dynamics in terms of (d) $x$-positions, (e) $y$-positions, (f) longitudinal velocities, $u$, and (g) transversal velocities, $v$.}
\end{figure*}

\noindent \textbf{A. Confined dynamics in narrow corridors.} Fig.~\ref{fig:narrow-corridor-trajectories} shows that our model can statistically reproduce single straight walking paths. We used a reduced spatial discretization $\Delta y=0.1$ m to capture small variations in the transversal direction. Fig.~\ref{fig:narrow-corridor-trajectories}c shows that the data-driven potential has a similar shape as the analytical potential proposed in~\cite{corbetta-pre-2017}. This demonstrates that the data-driven potential has an intuitive physical interpretation, underscoring that our method can be easily used to extract fundamental insights from the observed data.

\begin{figure*}
    \begin{minipage}{\columnwidth}
    \begin{overpic}[width=\columnwidth, trim = 0 -10 0 0]{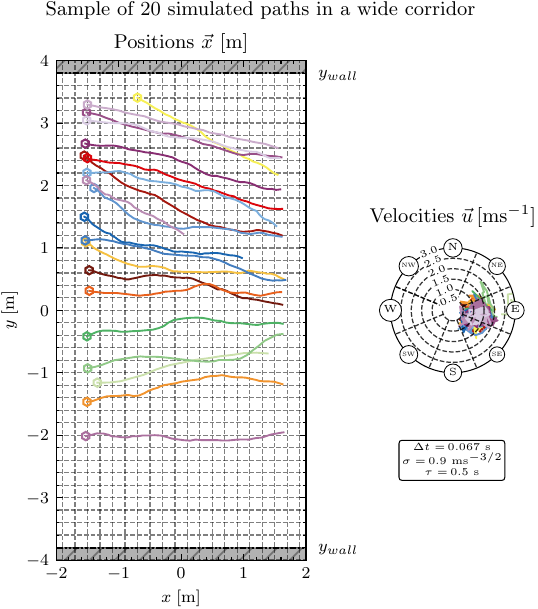}
        \put(-5,100){\Large B)}
        \put(0, 90){(a)}
        \put(58, 60){(b)}
    \end{overpic}
    \end{minipage}
    \begin{minipage}{\columnwidth}
    \begin{overpic}[width=\linewidth]{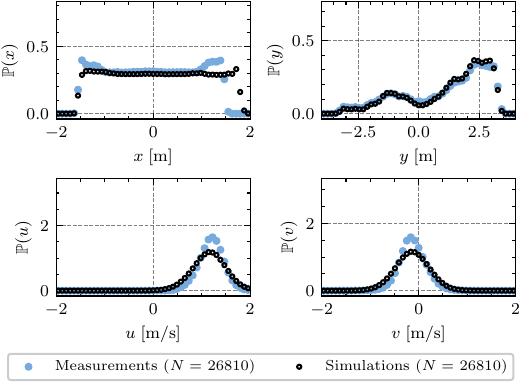}
        \put(0, 72){(c)}
        \put(52, 72){(d)}
        \put(0, 38){(e)}
        \put(52, 38){(f)}
    \end{overpic}
    \begin{overpic}[width=\linewidth, trim = 0 0 0 0]{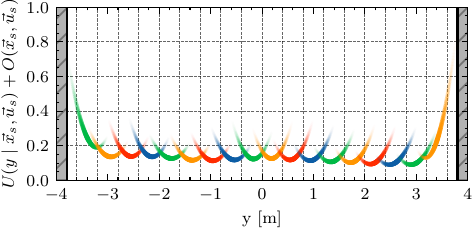}
        \put(-5,50){(g)}
    \end{overpic}
    \end{minipage}
    \caption{\label{fig:wide-corridor-trajectories} Modeling many parallel walking paths. (a,b) A sample with 20 simulated trajectories. (c -- f) A comparative analysis between the probability distributions of the recorded (blue dots) and simulated (open dots) trajectories across $x$-positions (c), $y$-positions (d), longitudinal velocity $u$ (e), and transversal velocity $v$ (f). The model  reproduces the statistics of the $x, y, u, v$. (g) Cross-section of the potential along the $x=0$ axis of the corridor, and conditioned to the following values of the slow variables: $x_s \approx 0$~m, $u_s \approx 0.7$~\ms, and $v_s \approx 0$~\ms. Each colored line segment represents a section of the potential conditioned to a different slow y-variable $y_s$. For visualization purposes we impose an offset $O(\vec{x}_s, \vec{u}_s) = A \ln{[\mathbb{P}(\vec{x}_s, \vec{u}_s)}]$.}
    \begin{overpic}[width=\columnwidth, trim= 0 -5 0 -5]{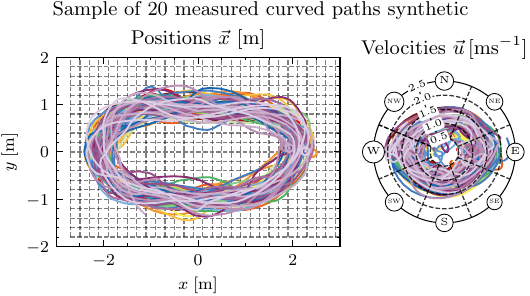}
        \put(-3,50){\Large C)}
        \put(0, 42){(a)}
        \put(67, 42){(b)}
    \end{overpic}
    \begin{overpic}[width=
    \columnwidth, trim= 0 -5 0 -5]{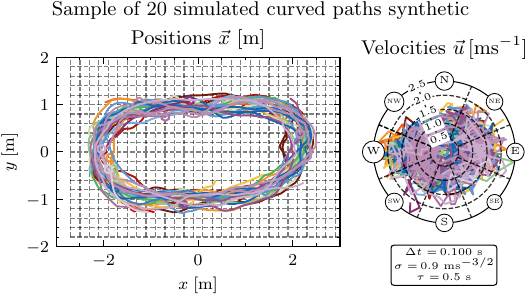}
        \put(0, 42){(c)}
        \put(67, 42){(d)}
    \end{overpic}
    \begin{overpic}[width=\histogramswidth]{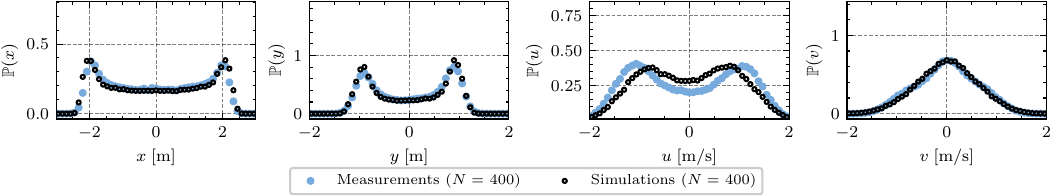}
        \put(0, 19){(e)}
        \put(26, 19){(f)}
        \put(51, 19){(g)}
        \put(76, 19){(h)}
    \end{overpic}
    \caption{\label{fig:curved-trajectories}Modeling curved walking paths. (a, b) A selection with 20 curved trajectories that we used as input to learn a curved potential. (c, d) Selection with 20 simulated trajectories using the model from Eq.~\eqref{eq:generalized-model-with-closure}. (e--h) A comparative analysis between the probability distributions of the recorded (blue dots) and simulated (open dots) trajectories across $x$, $y$, $u$, and $v$ in panels (e--h) respectively.}
\end{figure*}

\noindent \textbf{B. Paths in wide corridors: transversal translational symmetry broken.} 
Fig.~\ref{fig:wide-corridor-trajectories} shows that our model can statistically reproduce many parallel straight walking paths present in the setup of~\cite{corbetta-pre-2018}. Fig.~\ref{fig:wide-corridor-trajectories}g shows that the data-driven potential is even able to capture the continuous symmetry group of translations in the direction transversal to the corridor (cf. Sec.~\ref{ssec:wide-corridor}). 

\noindent \textbf{C. Fluctuations along curved paths.}
Fig.~\ref{fig:curved-trajectories} shows that our model can statistically reproduce a data set with curved, elliptical, paths. The data set is created using the model for curved paths reported by~\cite{vleuten-pre-2024}. Notably, our model allows us to significantly simplify the modeling construction in~\cite{vleuten-pre-2024} that leverages on a data-driven fit of local coordinate systems, which is used to identify Christoffel symbols.

\begin{figure*}
    \begin{overpic}[width=\columnwidth]{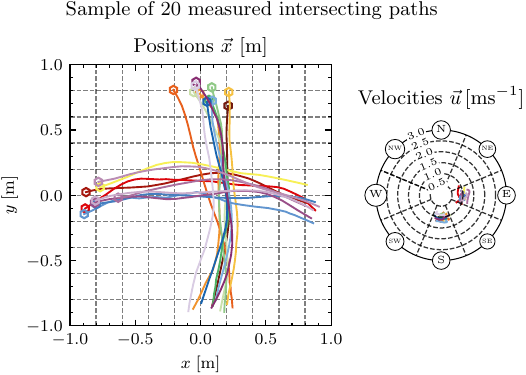}
        \put(-3,72){\Large D)}
        \put(0, 62){(a)}
        \put(67, 62){(b)}
    \end{overpic}
    \begin{overpic}[width=\columnwidth]{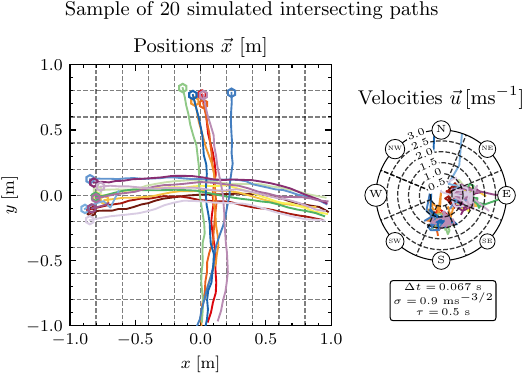}
        \put(0, 62){(c)}
        \put(67, 62){(d)}
    \end{overpic}
    \begin{overpic}[width=\histogramswidth, trim = 0 0 0 -10]{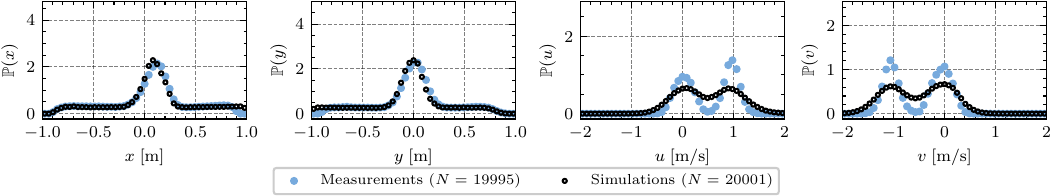}
        \put(0, 18){(e)}
        \put(26, 18){(f)}
        \put(51, 18){(g)}
        \put(76, 18){(h)}
    \end{overpic}
    \caption{\label{fig:intersecting-trajectories}Modeling intersecting walking paths. (a, b) We show 20 selected intersecting trajectories used as input. (c, d) 30 simulated trajectories using the generalized model Eq.~\eqref{eq:generalized-model-with-closure}. (e--h) A comparative analysis between the probability distributions of the measured (blue dots) and simulated (open dots) trajectories across $x$-positions (e), $y$-positions (f), $x$-velocities $u$ (g), and $y$-velocities $v$ (h).}
    \vspace{2em}
    \begin{overpic}[width=\columnwidth, trim = 0 -9 0 0]{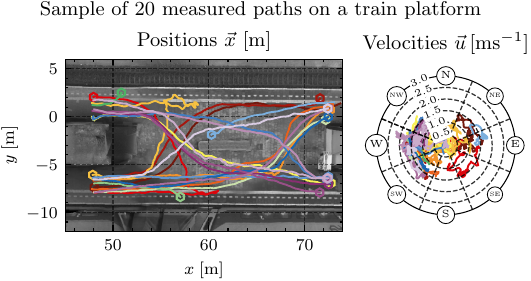}
        \put(-3,51){\Large E)}
        \put(0, 42){(a)}
        \put(67, 42){(b)}
    \end{overpic}
    \begin{overpic}[width=\columnwidth, trim = 0 -10 0 0]{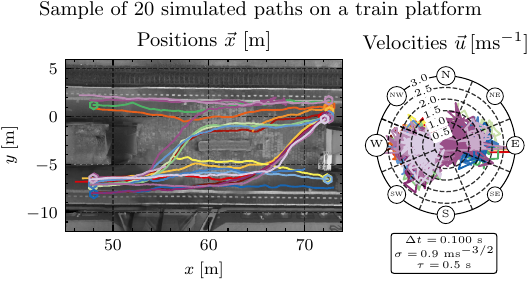}
        \put(0, 42){(c)}
        \put(67, 42){(d)}
    \end{overpic}
    \begin{overpic}[width=\histogramswidth]{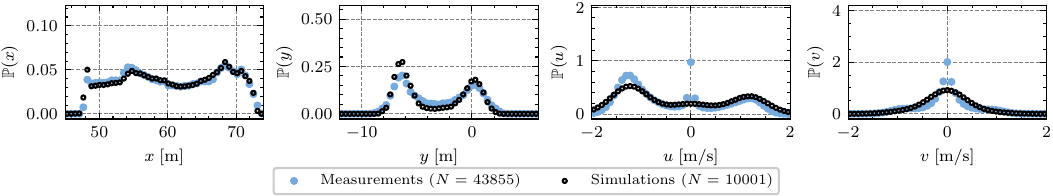}
        \put(0, 18){(e)}
        \put(26, 18){(f)}
        \put(51, 18){(g)}
        \put(76, 18){(h)}
    \end{overpic}
    \caption{\label{fig:station-trajectories}Modeling walking paths on a train platform at Eindhoven train station. (a, b) A selection with 20 measured trajectories. (c, d) A selection with 20 simulated trajectories. (e--h) A comparative analysis between the probability distributions of the measured (blue dots) and simulated (open dots) trajectories across x-coordinates $x$ (e), y-coordinates $y$ (f), longitudinal velocity $u$ (g), and transversal velocities $v$ (h). We observe a good resemblance between the spatial probability distributions and we note that all the dominant velocity modes are captured by the model.}
\end{figure*}

\noindent \textbf{D. Crosspoints and crossing trajectories.} Fig.~\ref{fig:intersecting-trajectories} shows that our model can statistically reproduce intersecting trajectories, which has not been done before. We test this on the basis of the narrow corridor trajectory data set (cf.~Fig.~\ref{fig:corridor-trajectories}(a)) for which we rotate half of the trajectories by 90 degrees, see Fig.~\ref{fig:intersecting-trajectories}a.

The richer symmetry structure of intersecting trajectories requires a description with further degrees of freedom including the tactical dynamics connected to the two different directions. This is addressed in our model by conditioning on the slow dynamics $\vec{x}_s$ and $\vec{u}_s$. 

\noindent \textbf{E. Physical insights in real-life contexts.}
Fig.~\ref{fig:station-trajectories} shows that our model can statistically characterize the complex dynamics of pedestrians walking across a train platform (cf.~\citep{vleuten-pre-2024} EHV data set). For simplicity, we assume a time-independent potential. Yet, the non-stationary dynamics in the train station due to the continuous arrival and departure of trains, could be easily incorporated in the model by learning a time-dependent potential. Additionally, our approach can be used to learn potentials associated with typical infrastructural elements, such as benches, staircases, doors, and corridors, allowing for testing modifications to existing environments or designing new ones.

\begin{table*}
\caption{\label{tab:validations} Pedestrian trajectory data sets for validation detailing the measurement location, a description of the dynamics, the number of recorded trajectories, the sampling frequency $f$, the noise intensity $\sigma$, the relaxation time $\tau$, the size of the lattice cells in terms of $\Delta x, \Delta y, \Delta v_r, \textrm{and} \, \Delta v_{\theta}$, relevant bibliographic references, and references to the figures with the modeling results.}
\begin{ruledtabular}
\begin{tabular}{l|p{37mm}|l|l|l|l|l|l|l|l|l|l|l}
 & Location & Description & Paths &  $f$ & $\sigma$ & $\tau$ & $\Delta x$ & $\Delta y$ & $\Delta v_r$ & $\Delta v_{\theta}$ & References & Results \\ 
  & & & [\#] & [Hz] & [m$\,$s$^{-3/2}]$ & [s] & [m] & [m] & [\ms] & [deg] & & \\
  \hline
A. & TU/e Metaforum & Narrow corridor & $2.0\cdot10^4$ & $15$ & 0.9 & 0.5 & 0.2 & 0.2 & 0.5 & 45 & \citep{corbetta-trp-2014, corbetta-pre-2017, corbetta-compscience-2019} & Fig.~\ref{fig:narrow-corridor-trajectories}\\
B. & Eindhoven railway station corridor & Wide corridor & $2.7\cdot10^5$ & $15$ & 0.9 & 0.5 & 0.2 & 0.2 & 0.5 & 45 & \citep{corbetta-pre-2018} & Fig.~\ref{fig:wide-corridor-trajectories} \\
C. & Synthesized with \citep{vleuten-pre-2024} & Curved paths & $3.7\cdot10^2$ & $10$ & 0.9 & 0.5 & 0.2 & 0.2 & 0.5 & 45 & \citep{vleuten-pre-2024} & Fig.~\ref{fig:curved-trajectories}\\
D. & Synthesized from TU/e Metaforum & Crosspoint & $2.0\cdot10^4$ & $15$ & 0.9 & 0.5 & 0.2 & 0.2 & 0.5 & 45 & \citep{corbetta-pre-2017, corbetta-compscience-2019} & Fig.~\ref{fig:intersecting-trajectories}\\
E. & Eindhoven railway station platforms 3 and 4 & Real-life env. & $4.4\cdot10^4$ & $10$ & 0.9 & 0.5 & 0.2 & 0.2 & 0.5 & 45 & \citep{vleuten-pre-2024} & Fig.~\ref{fig:station-trajectories}
\end{tabular}
\end{ruledtabular}
\end{table*}

\makeatletter\onecolumngrid@push\makeatother

\clearpage
\makeatletter\onecolumngrid@pop\makeatother
\section{Path-integral formulation}\label{sec:path-integral}
Previous works, e.g.~\citep{corbetta-compscience-2019}, have shown that the path-integral formalism, developed and commonly used in quantum mechanics \cite{feynman-book-1965,zinn-oxford-2021}, can be leveraged as an intuitive and trajectory-centric approach to describe pedestrian motion. This framework associates a probability with every possible path (impossible paths having zero probability). Although the Langevin formalism can be shown to be mathematically equivalent to the path-integral formalism \cite{zinn-oxford-2004,zinn-oxford-2021}, the latter provides an interesting and intuitive angle to understand the dynamics. This focus on whole trajectories may help to better explore fundamental questions such as: What are the most likely usage patterns? How are fluctuations around these paths characterized? What is the likelihood of rare and possibly dangerous events? How to make real-time forecasts of how a trajectory will continue?

\textbf{Probability of full paths}
Measurements of trajectories in real-world environments can be considered a statistical sampling from which we can infer the relative likelihood, $\mathcal{P}[\gamma]$, of observing any given path, $\gamma$, among the set of all possible paths. The probability density can be written as
\begin{equation}\label{eq: pi1}
    \mathcal{P}[\gamma]\,\mathcal{D}\gamma = \frac{1}{M}e^{-S[\gamma]} \, \mathcal{D}\gamma
\end{equation}
With $S[\gamma]$ the action functional that associates a scalar action $S$ to any path $\gamma$, and a normalization constant $M$. The expression above is the standard Wick-rotated quantum mechanical path integral, as ordinarily used in statistical field theory.

\textbf{Action functional} Understanding the action functional, $S[\gamma]$, is crucial as it fully characterizes all the properties of the motion and, if known, allows computation of all the statistical observables. However, this functional is not easily derivable from fundamental physical principles only. In this work, we have taken the first steps in deriving the action functional empirically from comprehensive trajectory data sets. We have shown that we can extract a potential that is capable of statistically reproducing the dynamics of pedestrian motion.

\textbf{Stationary paths} For the motion of a classical object, almost all the terms in the sum over paths cancel each other except the classical path. The stationary paths (usage modes) are defined as the trajectories for which the action is approximately equal for nearby paths (principle of stationary action), described in formulas as $S[x+\delta x] = S[x] + O(\delta x^2)$. 

So in the classical limit, a particle follows the trajectory that minimizes the action $S$. The principle of least (or stationary) action states that average trajectories $\gamma_m$, i.e. the most common paths, minimize the action $S[\gamma]$. For these trajectories, the variation in the action must go to zero $\Delta S[\gamma_m] = 0$. Traces the paths where the gradient to the potential is close to zero. A numerical sampling of the action functional enables us to simplify the complex dynamics recorded in millions of individual trajectories to the most relevant usage patterns. The path integral is dominated by the path with the least action. The knowledge of the action may easily allow to deduce the most commonly usage patterns (classical paths) as well as to estimate the probability associated to arbitrary path (e.g. real-time probability trajectory scoring).
\section{Conclusion}\label{sec:conclusion}
In this paper, we introduced a data-driven physics-based generalized Langevin model that allows robust and generic modeling of individual pedestrian behavior across generic environments hinging on extensive pedestrian trajectory data. Our model effectively captures the complex interplay between the deterministic movements and stochastic fluctuations associated with walking. A distinctive feature of our model is a conditioning on a pedestrian's tactical behavior -- i.e. their \textit{intended} walking path. Here we model this as a component of the dynamics happening on a slow manifold. Stochastic fluctuations, representing components such as sway, but also interactions with obstacles and, possibly, other pedestrians, perturb the trajectory by altering the path and velocity from their deterministic slow dynamics. Our approach generalizes models in the literature specifically tailored for narrow and wide corridors, which we reviewed in the second section of the paper.

The methods we presented can be easily applied to any pedestrian trajectory data set. Our data-driven framework first infers the slow dynamics from the actual measurements and then derives a piecewise approximation of the complex potential. We have successfully demonstrated the ability to completely learn a 1-person \textit{effective} (i.e. interactive) potential that depends on a person's actual dynamics and is conditioned on their slow dynamics. In this work, we assumed for the sake of simplicity a time-independent potential, however this can be easily generalized to carry additional dependencies, such as the local density and time. We focused on learning the potential from trajectory recordings and aimed to provide a simple framework that can be used to simulate pedestrian trajectories. For simplicity, we assumed the noise to be homogeneous in space. A logical next step would be to study the noise from the data and describe it in more detail.

Practical guidelines for selecting model parameters were established, reflecting normal walking conditions. We showed that our model can learn the dynamics by validating it across four complementary geometrically simple scenarios that are easy to understand. We demonstrated, through qualitative and quantitative comparisons -- including statistical analysis of the probability distributions -- that our model reliably reproduces realistic pedestrian dynamics. Additionally, we demonstrated our model's versatility by studying a complex, real-world, pedestrian environment to exemplify the model's ability to disclose valuable insights from large-scale human trajectory recordings.

The novel tools developed in this study provide valuable assets for urban planning and the design of public spaces. Reading the potential fields gives a higher level of physical understanding of the environment. On one hand, synthesized potentials associated with typical infrastructural elements -- e.g. benches, staircases, doors, and corridors -- enable a modular approach, facilitating the combination of these building blocks into entirely new environments. On the other hand, analyzing specific, time-dependent, potentials across various crowd densities can provide fundamental insights about individual and collective behaviors. 
By situating our results within the path-integral framework, we emphasize its theoretical value as a trajectory-centric framework to describe any possible dynamics. Parallel to the principle of stationary action we demonstrated a method to simplify complex dynamics, recorded in millions of trajectories, to their classical paths. Extending its application well beyond the context of pedestrian dynamics.

\section*{Supplementary material}
We provide an open-source Python implementation of the generalized data-driven pedestrian model on GitHub: \href{https://github.com/c-pouw/physics-based-pedestrian-modeling}{physics-based-pedestrian-modeling} \cite{pouw-physped-2024}. The code can be freely used to create models for any extensive trajectory data set. 

\begin{acknowledgments}
This work is part of the HTSM research program “HTCrowd: a high-tech platform for human crowd flows monitoring, modeling and nudging” with project number 17962, financed by the Dutch Research Council (NWO).
\end{acknowledgments}

\appendix
\section{\label{app:scaling-factors}Scaling factors}
We extract a potential using 
\begin{equation}
    \prob(\vec{x}, \vec{u} \mid \vec{x}_s, \vec{u}_s) \propto e^{-U(\vec{x}, \vec{u} \mid \vec{x}_s, \vec{u}_s)}.
\end{equation}

We consider a behavior that is locally equivalent to that of an harmonic oscillator with active friction
\begin{align}
    U_x(x) &= \beta_x\left(x-\mu_x\right)^2 = \beta_x x'^2 \\
    U_u(u) &= \beta_u\left(u - \mu_u\right)^2 = \beta_u u'^2 \\
    U(x,u) &= U_x(x)+ U_u(u) = \beta_x x'^2 + \beta_u u'^2,
\end{align}
From which 
\begin{align}
    -\log(\prob_{exp}(u)) &= \frac{2 \beta_u}{\sigma^2}u'^2 + K' \label{eq:A5}\\
    -\log(\prob_{exp}(x)) &= \frac{4 \beta_x \beta_u}{\sigma^2}x'^2 + K'' \label{eq:A6}.
\end{align}
Given the definition of $\prob_{exp} (u) \propto \exp{(-u'^2/2\, \xi_u^2)}$ and Eq.~\eqref{eq:A5} we derive

\begin{align}
    \frac{2 \beta_u u'^2}{\sigma^2}  &= \frac{u'^2}{2 \, \xi_u^2} \\
    \beta_u &= \frac{\sigma^2}{4 \xi_u^2},
\end{align}
and similarly with $\prob_{exp}(x) \propto \exp{(-x'^2/2 \, \xi_x^2)}$ and Eq.~\eqref{eq:A6} we derive
\begin{align}
    \frac{4 \beta_u \beta_x x'^2}{\sigma^2} &= \frac{x'^2}{2 \, \xi_x^2} \\
    \beta_x &= \frac{\sigma^2}{8 \beta_u \xi_x^2} = \frac{\xi_u^2}{2 \, \xi_x^2}
\end{align}

\section{\label{app:numerical-simulations}Numerical simulations}
We initialize new simulations with a starting position $\vec x(t=0)$ and starting velocity $\vec u(t=0)$, through a sampling of the origins observed in the measurements. The simulation is advanced in time by integrating the stochastic differential equations leveraging an order 1.0 strong stochastic Runge-Kutta method \citep{rossler-siam-2010}. Simulations are terminated when they reach a state with infinite potential or after 
a maximum simulation time $t_{max}$.

\bibliographystyle{apsrev4-2}
\bibliography{references.bib}

\end{document}